\documentclass{article}
\usepackage{amsfonts}
\usepackage{esvect}

\newcommand{\vctr}[1]{\ensuremath{\mathbf{ #1 }}}

\newcommand{\dr}[1]{\ensuremath{\mathrm{d} #1\,}}
\newcommand{\mc}[1]{\ensuremath{\mathcal{#1}}}

\newcommand{\be}{\begin{equation}}
\newcommand{\ee}{\end{equation}}
\newcommand{\e}[1]{\mathrm{e}^{#1}}

\newcommand{\iec}{\mbox{i.\,e.\,}}

\newcommand{\egc}{\mbox{e.\,g.\,}}

\usepackage{chicago}

\renewcommand{\Re}{\mathbb{R}}

\newcommand{\Co}{\mathbb{C}}
\renewcommand{\vctr}[1]{\vv{\mathbf{#1}}}

\begin{document}
\title{Gauge Invariance through Gauge Fixing}
\author{David Wallace\thanks{Department of History and Philosophy of Science / Department of Philosophy, University of Pittsburgh, PA 15260, USA; email \texttt{david.wallace@pitt.edu}}}
\maketitle

\begin{abstract}
Phenomena in gauge theory are often described in the physics literature via a specific choice of gauge. In foundational and philosophical discussions this is often criticized as introducing gauge dependence, and contrasted against (often aspirational) ``gauge-invariant'' descriptions of the physics. I argue, largely in the context of scalar electrodynamics, that this is misguided, and that descriptions of a physical process within a specific gauge are in fact gauge-invariant descriptions. However, most of them are \emph{non-local} descriptions of that physics, and I suggest that this ought to be the real objection to such descriptions. I explore the unitary gauge as the exception to this nonlocality and consider its strengths and limitations, as well as (more briefly) its extension beyond scalar electrodynamics.
\end{abstract}

\section{Introduction: who's afraid of gauge fixing?}\label{introduction}

Consider the theory of a complex scalar field interacting electromagnetically (`scalar electrodynamics', as particle theorists call it). In this theory, the fields are represented by ordered pairs ($\psi,A$), where $\psi$ is a complex scalar field on Minkowski spacetime on \mc{M} and $A$ is a covariant vector field (\iec, one-form field). We normally also include some boundary condition: asymptotically at spatial infinity, $A$ tends to zero and $\psi$ tends to some constant value.\footnote{In some applications of gauge theory --- such as $SU(2)$ gauge theory in three spatial dimensions --- it is possible to have a spatially varying boundary condition (see, \egc, \cite[section 15.3]{chengli}; I set this aside for simplicity.} Theories like this are varyingly used to describe classical field theories, the theory of a quantum particle moving in a background classical electromagnetic field, and (normally through the mechanism of the path integral, and/or using the formalism of the quantum effective action) scalar electrodynamics, the theory of a complex quantum scalar field interacting with a quantum electromagnetic field. I write $\mc{F}$ for the space of all such fields satisfying the boundary conditions and whatever smoothness requirements a particular theory leads us to impose.

It is key to theories of this kind that their dynamics (however formulated) is \emph{gauge-invariant}: invariant, that is, under transformations of the form
\be
\psi(x) \rightarrow U(x)\psi(x),\,\,\,\,\,\, A_\mu(x) \rightarrow A_\mu(x) - \frac{1}{e}U^{-1}(x) \partial_\mu U(x)
\ee
where $U$ is a smooth complex function on spacetime satisfying $|U(x)|=1$, or equivalently (since Minkowski spacetime is simply connected),
\be
\psi(x) \rightarrow \e{ie\theta(x)}\psi(x),\,\,\,\,\,\, A_\mu(x) \rightarrow A_\mu(x) - \partial_\mu \theta(x)
\ee
where $\theta$ is a smooth real function on spacetime. (In either case, $e$ is a nonzero real number representing the strength of the electromagnetic coupling.) These transformations are called \emph{local} $U(1)$ transformations, since $U$ can also be thought of as a function from spacetime to the group $U(1)$; the group of all such transformations is $\mc{G}$.

This framework readily generalizes to \emph{non-Abelian gauge theory}, where now $\psi$ is not just a complex field but takes values in some space $\mc{V}$, and where $A$ is a one-form taking values in a Lie algebra of operators on $\mc{V}$; it also generalizes to theories where $\psi$ is a spinor or vector field. In each case there is a generalization of the notion of a gauge group, with $U$ taking values in a group larger than $U(1)$ and applying a somewhat more complicated transformation rule to $A$. In this general framework, if $G$ is the group in which $U$ takes values, the gauge group $\mc{G}$ is the group of all such $U$, \iec all smooth maps from spacetime to $G$. The technical details are well known (see, \egc, \cite[ch.15]{weinbergqft2}) but will not be needed here. (It is also common in the philosophical literature --- see, e.g., (\citeNP{healeybook,weatherallgauge}) --- to use the more elegant, but more abstract, framework of fiber bundles, so that $\psi$ becomes a section of a $\mc{V}$-bundle over spacetime, $A$ becomes a connection on that bundle, and \mc{G} becomes a group of automorphisms of that bundle but for my purposes this is unnecessary.)

Gauge symmetry in physics is normally taken as a sign of \emph{representational redundancy}: the pair ($\psi,A$) overdescribes the physical goings on, so that distinct such pairs actually describe the same physics. To be more precise: let's say that a gauge transformation is \emph{small} if its defining function $U$ tends to the identity at infinity and if it can be smoothly deformed to the identity while preserving this boundary condition. (For $U(1)$ gauge theory on Minkowski spacetime, this second condition is actually redundant, but for some gauge theories, and some background topologies, there are gauge transformations that equal the identity at the boundary but cannot be deformed to the identity everywhere while respecting the boundary condition.) And a transformation is \emph{boundary-preserving} if it leaves the spatial boundary condition invariant. Any small gauge transformation is boundary-preserving, but the converse need not be true. Small and boundary-preserving transformations each form a group, which I write respectively as $\mc{G}_S$ and $\mc{G}_{BP}$.

Then the normal interpretation of gauge transformations in the physics literature is:
\begin{itemize}
\item Small gauge transformations simply encode representational redundancy: fields related by a small gauge transformation are the same field differently described. The group $\mc{G}_S\subset \mc{G}$ of small gauge transformations is not a \emph{physical} symmetry group at all, but just a representation of that redundancy. 
\item Large gauge transformations, insofar as they are boundary-preserving (that is, insofar as they are defined at all as transformations on the space of fields) are physically-meaningful symmetry transformations in the same way as (say) the global $U(1)$ symmetries of the Klein-Gordon field, or the translation and rotation symmetries of Euclidean space. The quotient group $\mc{G}_{BP}/\mc{G}_S$ (whose action on a classical field is defined up to a small gauge transformation) is the physical (internal) symmetry group of the theory.
\end{itemize}
(I defend this physics orthodoxy \emph{in extenso} in Wallace~\citeyear{wallace-isolated-1,wallace-isolated-2,wallaceobservingsymmetries}; see also \cite{belotelvis}. Here I just assume it.)

At a somewhat abstract level, the true space of classical fields is not $\mc{F}$ but the quotient space $\mc{F}/\mc{G}_S$, whose elements are equivalence classes under the equivalence relation $(\psi,A)\sim (\psi',A')$, which holds between two field pairs iff they are related by a small gauge transformation. But this space is hard either to calculate with concretely, or to interpret physically. In mainstream physics, the normal route to address this problem is \emph{gauge fixing}: we come up with some condition on field pairs $(\psi,A)$ such as to pick out exactly one element of each equivalence class. 
It will be helpful to have some specific examples\footnote{For details, see any modern quantum field theory textbook, \egc  \cite{weinbergqft2,peskinschroeder}.} (in each case from scalar electrodynamics):
\begin{description}
\item[Coulomb gauge:] Choose a space/time separation on Minkowski spacetime, write $A=(V,\vctr{A})$, and impose the condition that $\vctr{\nabla}\cdot \vctr{A}=0$. To see that this does successfully define a gauge fixing, consider an arbitrary field $(\psi,V,\vctr{A})$ and look for the gauge transformation $(\psi,V,\vctr{A})\rightarrow (\e{ie\theta}\psi,V-\dot{\theta},\vctr{A}-\vctr{\nabla}\theta)$ that puts it into Coulomb gauge. We can easily see that $\theta$ solves
\be
\nabla^2 \theta = \vctr{\nabla}\cdot \vctr{A}
\ee
which has a guaranteed solution that is unique given the boundary conditions. So the gauge fixing condition picks out \emph{at least} one element of each equivalence class (because some such $\theta$ can always be found) but also \emph{at most} one element (because $\theta$ is unique). Note that Coulomb gauge is purely a condition on the vector-field part of the field pair and does not constrain the matter field.
\item[Lorenz gauge:] require $\partial_\mu A^\mu=0$. By essentially the same argument as for Coulomb gauge, any field is related to a field in Lorenz gauge by a gauge transformation unique up to some function $\chi$ satisfying $\partial_\mu\partial^\mu \chi=0$. Given some boundary conditions at infinity, this makes Lorenz gauge a valid gauge transformation provided that it is supplemented by some additional gauge-fixing term on the initial (or final) data (e.g. that it satisfies $\partial_i A^i=0$ at the time). Lorenz gauge is again a condition purely on the vector field.
\item[Unitary gauge:] require that $\psi$ is real and positive. Since we can write $\psi=\rho \e{i\phi}$, it is immediate for nonzero $\rho$ that the gauge transformation $\theta = \phi/e$ uniquely transforms a field into unitary gauge.  Unitary gauge, unlike our first two examples, is a condition purely on the scalar field, leaving the vector field unconstrained. (Unitary gauge has some subtleties, especially in the case where $\rho=0$, which I consider in more detail later.)
\item [$R_\xi$ gauge:] Writing $\psi(x)=(\rho_0+v + i b)$ for some nonnegative real number $\rho_0$ (usually whatever the boundary condition is) we impose
\be
\partial_\mu A^\mu=e \xi v b
\ee
for arbitrary $\xi$. $R_\xi$ gauge is a hybrid of unitary and Lorenz gauge, and places a joint constraint on scalar and vector fields. (I leave it to the reader to demonstrate that it successfully fixes a gauge, given appropriate initial and boundary conditions.)
\end{description}
The normal assumption in physics is that gauge fixing is necessary to remove the descriptive redundancy, but that the specific choice of gauge fixing is a situation-dependent matter of convenience and pragmatics. The Coulomb gauge is excellently suited to reveal the degrees of freedom of the electromagnetic field, but is awkward calculationally because it breaks Lorentz invariance; Lorenz gauge is much more convenient in that respect. Unitary gauge is conceptually helpful in understanding the Higgs mechanism, but $R_\xi$ gauge is better suited to addressing renormalization in Higgs calculations.\footnote{See, \egc, \cite[section 21.2]{weinbergqft2} for a discussion of this point.} A choice of gauge, just like a choice of coordinate system, is a necessary preliminary to doing physics.

In the foundational literature, and in philosophy of physics, the mainstream physics approach has been met with suspicion. The general concern\footnote{An important exception is \citeN{struyvegauge}, who notes (p.230) that `if one wants to take seriously the ontology suggested by gauge fixing, then there is very little difference compared to an ontology in terms of gauge independent variables.} is that since only the gauge-invariant features of the field are physically meaningful, any gauge-fixed description mixes up genuinely gauge-invariant (and so, real) features of the physical problem with gauge-dependent (and so, illusory) features that just reflect our conventional choice of one gauge-fixing over another. The main foci of this literature are the Higgs mechanism and the Aharonov-Bohm (A-B) effect. In the former case, the definitive statement of the objection remains John Earman's (\citeyearNP[189-190]{earmancurie}):
\begin{quote}
The popular presentations use the slogan that the vector field
has acquired its mass by ``eating'' the Higgs field. But, as the authors of the
standard treatises well know but rarely bother to warn the unwary reader,
talk of \emph{the} Higgs field has to be carefully qualified since by itself, the value
of $[\psi]$ does not have gauge invariant significance. The popular slogan can be
counterbalanced by the cautionary slogan that neither mass nor any other
genuine attribute can be gained by eating descriptive fluff.\ldots it is dereliction of duty for philosophers to repeat the
physicists' slogans rather than asking what is the content of the reality that
lies behind the veil of gauge.
\end{quote}
In the latter, one of the early influential statements is Healey's~(\citeyearNP{healey1995}, p.22):
\begin{quote}
[T]he A-B effect is local only if [the vector potential] is a physically 
real field, capable of acting on the electrons directly. But there is reason 
to doubt that the magnetic vector potential is a physically real field, 
since [it] is not gauge-invariant, unlike the magnetic field $\vctr{B}$ and the 
phase-shift\ldots
\end{quote}

The search for gauge-invariant accounts of the A-B effect is worth reviewing in more detail, as a reference point for later considerations. The usual argument goes: in classical electromagnetism the field strength
\be F_{\mu\nu}=\partial_\nu A_\mu - \partial_\mu A_\nu\ee 
(or its electric and magnetic components 
\be \vctr{E}=\dot{\vctr{A}}-\vctr{\nabla}V;\,\,\,\,\vctr{B}=\vctr{\nabla} \times \vctr{A}\ee
in any given inertial frame) are gauge-invariant and provide a complete description of the gauge-invariant features of the theory, but this ceases to be true when we consider the interaction of an electric field with a complex charged field, whether understood classically or through quantum particle mechanics or QFT. For consider a static magnetic field defined on a space coordinatized with Lorentz-frame coordinates $(t,x,y,z)$ from which the $z$-axis has been removed from space at each instant in time (this `solenoid spacetime' might be used to model an idealized solenoid). We take the electromagnetic field $A=(V,\vctr{A})$, in Coulomb gauge, to be
\be\label{solenoid}
V(t,z,y,z)=0;\,\,\,\,\vctr{A}(x,y,z)=\frac{\Phi}{2\pi (x^2+y^2)}(y,-x,0)
\ee
We can readily calculate that $\vctr{E}=\vctr{B}=0$, but the quantity
\be
\Delta = \oint_C \vctr{A}\cdot \dr{\vctr{x}} = \Phi
\ee
(where $C$ is any path enclosing the missing $z$-axis) is equally readily demonstrated to be gauge-invariant.\footnote{More accurately, it is invariant under \emph{small} gauge transformations; large gauge transformations can modify it by integer quantities.} And this quantity is physically significant: $\Delta e$ is exactly the induced phase shift for interference between beams of charged particles that pass to either side of the solenoid. So we have gauge-inequivalent $A$-fields, but that inequivalence is invisible as long as we have only the field strengths. If we de-idealize the problem by restoring a physical solenoid along the $z$-axis, we need to thread a magnetic field of total flux $\Phi$ (strictly speaking, $\Phi+n/e$) through a narrow tube around the axis, but the magnetic field remains zero in the region through which the charged particles actually pass. So we seem to be left with a trilemma (explored in detail by \citeN{healeybook}):
\begin{enumerate}
\item We could take the field to be characterized by the field strengths alone --- but then that characterization is inadequate for non-simply-connected spacetimes, and commits us to a form of action at a distance even in Minkowski spacetime.\footnote{The field strengths also obey constraint equations which arguably should not be seen as dynamical but as encoding nonlocal structure; thanks to an anonymous referee for this observation.}
\item We could take it to be characterized by the holonomies
\be
H[C]=\oint_C A_\mu\cdot \dr{x^\mu}
\ee
where now $C$ is any closed path in spacetime --- but now our characterization (as well as being extremely awkward calculationally) is non-separable, since the separate holonomies in regions $A$ and $B$ will underdetermine the holonomies in $A\cup B$ for many choices of $A$ and $B$ (most obviously, for any decomposition of a non-simply-connected set into simply connected subsets).
\item Or we could just take it to be characterized by the field $A$ --- but now that characterization is gauge-dependent, and (so long as we hold on to the idea that gauge-related fields are just different representations of the same physical goings-on) our description obscures matters by mixing up genuine physics and descriptive fluff.
\end{enumerate}

I pause to notice an oddity about this (widely used) approach to the A-B effect. It assumes \emph{vacuum} electrodynamics, or else electrodynamics against a background of externally-imposed charge and current densities, so that we have to characterize only the electromagnetic field, and can neglect the charged matter field. But the equations of classical electrodynamics are
\be
\partial_\nu F^{\mu\nu}=4\pi J^\mu
\ee
where the current $J^\mu$ is either stipulated as background or dynamically generated from some gauge-invariant phenomenological model, such as charged dust. And the (internal) dynamical symmetry group of this theory is not just local $U(1)$ symmetries, but the wider group of symmetries
\be\label{widergroup}
A \rightarrow A + \chi
\ee
where $\chi$ is any closed one-form field: that is, any one-form field satisfying $\partial_\mu \chi_\nu+\partial_\nu \chi_\mu$. (On a simply-connected manifold a one-form is closed if and only if it is exact (can be expressed as the derivative of a function) but this is not the case on multiply-connected manifolds, like our idealized-solenoid spacetime.)  Holonomies are not invariant under these transformations; indeed, it is easy to show that the $A$-field on a region is fully characterized, up to transformations of form (\ref{widergroup}), by the field strength $F$ on that same region, so that after all the field strength provides a local, gauge-invariant characterization. It is exactly because electromagnetism is invariant under this wider symmetry group that the Aharonov-Bohm effect requires quantum mechanics (or at any rate coupling to a complex field that also transforms under local $U(1)$). But in that context, we should be looking for a gauge-invariant description not just of $A$ but of the pair $(\phi,A)$ --- and neither the field strength nor the holonomy does that. I expand on this observation in \cite{wallaceAB}.

I return to the issue of characterizing $(\psi,A)$ and not just $A$ in section \ref{locality}. For now I want to point out a more basic problem with the idea that we need to replace gauge-fixed accounts of electromagnetic physics with genuinely gauge-invariant ones. Consider again the gauge-fixed expression (\ref{solenoid}) for the solenoid spacetime's scalar and vector potential in the Coulomb gauge, and pick a specific $t,x,y,z$. The property that an electromagnetic potential has if, in the Coulomb gauge, it has value $(\Phi/2\pi(x^2+y^2))(0,y,-x,z)$ at point $(t,x,y,z)$ \emph{is a gauge invariant property}.

How can this be? Well: a gauge-invariant property is one that holds of a field $(\psi,A)$ iff it holds of any gauge-related field $(u_*(\psi,A))$. The statement that the $A$-field at $(t,x,y,z)$ has such-and-such value, \emph{simpliciter}, is obviously not gauge-invariant: any gauge transformation with nonzero derivative at $(t,x,y,z)$ will change it. But two fields that are gauge related, are, by definition, exactly the same when expressed in a specific gauge. 

We can go further. The statement ``there is a gauge in which the $A$-field is given by (\ref{solenoid})'' is also gauge-invariant. It holds of a field $A$ iff there is a gauge transformation that takes it to (\ref{solenoid}). But of course if $A$ and $A'$ are gauge-related then this holds of one of them iff it holds of the other.

(There is a close analogy to the equally-counter-intuitive fact that we can use coordinate systems to give coordinate-independent specifications of the distribution of matter. One way to say that a given curve in $\Re^2$ is diffeomorphic to the circle, for instance, is to say that there is a smooth coordinatization of $\Re^2$ in which the curve satisfies $x^2+y^2=1$. I expand on this feature of coordinate systems in \cite{wallacecoordinates}.)

So: gauge-fixing is not an alternative to a gauge-invariant characterization of the fields of scalar electrodynamics; it is a way to provide a gauge-invariant characterization. And insofar as there is reason to find such characterizations unsatisfactory, that reason cannot be that they are gauge-variant.

Let us see what it might be.

\section{A framework for gauge-invariant descriptions}\label{framework}

It will be helpful to consider the problem in a more general and abstract framework. Let's say that a \emph{gauge field theory} is a triple $( \mc{M},\mc{F},\mc{G}_S)$, where
\begin{itemize}
\item $\mc{M}$ is a manifold (possibly with additional structure);
\item $\mc{F}$ is a space of fields on $\mc{M}$. I leave intentionally vague exactly what `field' means here: certainly I intend it to cover at least the usual collections of scalar, vector, tensor, and spinor fields one finds in particle physics. 
But I require fields to satisfy two conditions:
\begin{enumerate}
\item[(i)] \emph{Fields are local:} if $\varphi$ is a field on $\mc{M}$, it has a well-defined restriction to a field $\varphi|_X$ on any open subset $X\subset \mc{M}$, satisfying the consistency requirement that $(\varphi|_X)|_Y=\varphi|_Y$ for $Y\subset X\subset \mc{M}$.
\item[(ii)] \emph{Fields are separable:} If $\{X_i\}$ is a collection of open subsets of $\mc{M}$ with $\cup_iX_i=X$, then for any two fields $\varphi,\varphi'$, if $\varphi|_{X_i}=\varphi'|_{X_i}$ for each $i$ then $\varphi|_X=\varphi'|_X$. (In other words, a field is specified in a region by specifying its values in elements of an arbitrarily small partition of the region.)
\end{enumerate}
$\mc{F}_X$ is the space of fields restricted to $X$.
\item $\mc{G}_S$ is a group of bijections of $\mc{F}$ --- the (small) gauge transformations of the gauge field theory. Any element $g\in \mc{G}_S$ has a restriction $g|_X$ to any open subset $X\subset \mc{M}$, and this restriction commutes with field restriction: $(g|_X)_*(\varphi|_X)=(g_*\varphi)|_X$.  Two restricted fields $\varphi,\varphi'$ on $X$ are gauge-equivalent iff there is some $g\in \mc{G}_S$ such that $g|_A(\varphi)=g|_A(\varphi')$; as a special case, two fields on the whole manifold are gauge-equivalent iff they are transformed into one another by some $g\in \mc{G}_S$.
\end{itemize}
Of course, this is a very attenuated notion of physical field, lacking any notion of dynamics, but it will suffice for our purposes. Scalar electrodynamics clearly fits this framework: the fields are pairs $(\varphi,A)$ satisfying the boundary conditions, the small gauge transformations are the small local $U(1)$ transformations. So do non-Abelian gauge theories, and theories with spinor or tensor matter fields.

Given this setup, a \emph{descriptor} $d$ for the theory is a map $d:\mc{F}\rightarrow \mc{D}$, where $\mc{D}$ is any space we like. A descriptor might return a field's value at some spacetime point, or its average over some spacetime region, or its asymptotic behavior, or it might be a map taking values in $\{T,F\}$, returning $T$ iff the field  has some specified property.

A descriptor $d$ is \emph{gauge-invariant} if $d(\varphi')=d(\varphi)$ whenever $\varphi'$ and $\varphi$ are gauge-equivalent, \iec whenever there is some $g\in \mc{G}_S$ such that $g_*\varphi=\varphi'$. Conversely, a descriptor is 
\emph{complete} if whenever $d(\varphi)=d(\varphi')$, $\varphi$ and $\varphi'$ are gauge-equivalent. A complete gauge-invariant descriptor provides a non-redundant, gauge-invariant description of the field theory: if we know that a field is mapped to some $x\in \mc{D}$, we know which field it is up to gauge invariance.

In electrodynamics, each value $F(x)$ of the field strength at spacetime point $x$ is a descriptor; so is the overall field $F$. Both are gauge-invariant; the latter is a complete descriptor for vacuum electrodynamics provided that the spacetime is simply connected. Each holonomy $\Delta(C)$ is a (gauge-invariant) descriptor; so is the holonomy map $\mc{C}\rightarrow \Delta (\mc{C})$; the latter is complete for vacuum electrodynamics whatever the topology of the spacetime. The vector field itself at a point, $A(x)$, is a descriptor; so is the matter field $\psi(x)$; so is the (trivial) descriptor that just assigns each field pair $(\psi,A)$ to itself. The latter is complete; neither are gauge-invariant.

Gauge fixings are also easily described in this framework: a gauge-fixing $f$ is a map from $\mc{F}$ to itself with the properties that (i) $f(\varphi)$ is gauge-equivalent to $\varphi$, and (ii) if $\varphi$ and $\varphi'$ are gauge-equivalent, $f(\varphi)=f(\varphi')$. (This is just to say that a gauge-fixing picks out a preferred member of each gauge equivalence class.

My point in the previous section is now simply that \emph{gauge fixings are complete gauge-invariant descriptors}. This is easy to demonstrate, indeed almost immediate. Recall that a gauge fixing selects one element from each class of gauge-equivalent fields. So it can be represented by a map $f$ from $\mc{F}$ to itself, such that $f(\varphi)=f(\varphi')$ iff $\varphi$ and $\varphi'$ are gauge-equivalent. But that is just the condition for $f$ to be a complete gauge-invariant descriptor.

So what --- if anything --- is wrong with just saying that we extract a theory's gauge-invariant content just by choosing a gauge?

\section{Locality}\label{locality}

To answer this, let's specialize again: back to scalar electrodynamics, and specifically to Coulomb gauge. We have
\be
\vctr{\nabla} \times (\vctr{\nabla}\times \vctr{A})\equiv \vctr{\nabla}\times \vctr{B};
\ee
using the identity $\vctr{\nabla}\times \vctr{\nabla}\times \vctr{A}=\vctr{\nabla}(\vctr{\nabla} \cdot \vctr{A}) - \nabla^2 \vctr{A}$ and the Coulomb gauge condition, we get
\be
\nabla^2 \vctr{A}=\vctr{\nabla}\times \vctr{B}.
\ee
The right side of this equation is gauge-invariant, and (given boundary conditions) the equation can be solved explicitly to give the left side in terms of the right:
\be\label{coulombcalc}
\vctr{A}(\vctr{x},t)=\frac{1}{4\pi}\int\dr{x'^3}\frac{\vctr{\nabla} \times \vctr{B}(\vctr{x'})}{|\vctr{x}-\vctr{x}'|} 
\ee
On the one hand, this (combined with the equivalent result for $V$) is an explicit definition of my general observation that a gauge-fixed potential is uniquely determined by gauge-invariant features of the field, and so is itself gauge-invariant. 

On the other hand, it is equally true that the gauge-fixed potential is \emph{non-locally} dependent on manifestly local gauge-invariant features of the field like $\vctr{B}$.  To determine $\vctr{A}(t,\vctr{x})$ in the Coulomb gauge we need to know the magnetic field arbitrarily far from $\vctr{x}$ (albeit the degree of dependence falls off with the inverse square of distance). $\vctr{A}(t,\vctr{x})$ may be a gauge-invariant quantity, but it is not in any physically meaningful sense localized at $(t,\vctr{x})$, despite its deceptive formulation as a vector field.

With this in mind, let's return to our abstract formalism and ask how to think about locality. We can define a \emph{local descriptor} $d$ as follows: it is specified (for a gauge field theory $(\mc{M},\mc{F},\mc{G}_S)$) by:
\begin{itemize}
\item A map $X\rightarrow D_X$ that assigns a (possibly structured) space $D_X$ to each open subset $X\subset\mc{M}$ 
\item For each pair of open subsets $X,Y$ of \mc{M} with $Y\subset X$, a restriction map $x\rightarrow x|_Y$ from $D_X$ to $D_Y$, such that $(x|_Y)|_Z)=x|_Z$ (hence the notation $x|_Y$ is consistent without needing to mention what $D_X$ $x$ belongs to). 
\item For each open $X\subset \mc{M}$, a map $d_X:\mc{F}_X\rightarrow D_X$, which commutes with restriction: 
\be
(d_X(\varphi))|_Y = d_Y(\varphi|_Y).
\ee
\end{itemize}
The idea is that $d$ does not just describe the entire field: for any open $X$, it describes the field \emph{on X}. Clearly any local descriptor is a descriptor.

We define a local descriptor as:
\begin{itemize}
\item $\emph{separable}$ iff, given any collection of sets $\{X_i\}$ with $\cup_i X_i=X$, if $d_{X_i}(\varphi|_{X_i})=d_{X_i}(\varphi'|_{X_i})$ then $d_X(\varphi|_X)=d_X(\varphi'|_X)$. In other words, a local descriptor is separable iff its description of a field on $X$ depends only on its descriptions of the restrictions of the field to arbitrarily small elements of a partition of $X$.
\item \emph{Locally complete} iff for any $X\subset \mc{M}$, and any fields $\varphi,\varphi'\in \mc{F}_X$, if $d_X(\varphi)=d_X(\varphi')$ then $\varphi$ and $\varphi'$ are gauge-equivalent.
\end{itemize}

We can apply this framework to the usual gauge-invariant descriptors used in discussions of the A-B effect. The map $A\rightarrow F$ is a local descriptor, assigning to each gauge potential on a region its field strength on that same region; it is separable, but not locally complete, even though it is complete (as long as we continue to ignore matter fields) on simply connected manifolds. That is: knowing $F$ everywhere on a simply connected space fixes the connection on that space up to gauge equivalence, but knowing $F$ on a multiply-connected subset of that space does not fix the connection on that subset, even up to gauge equivalence. Conversely, the assignment to each $X\subset \mc{M}$ of the holonomies of curves in $X$ (which is also a local descriptor) is  locally complete, but not separable.

As for gauge-fixings: the Coulomb gauge is not a local descriptor at all, despite its deceptively field-theoretic format: we cannot gauge-transform the restriction of a field to $X\subset \mc{M}$ into Coulomb gauge without knowing the gauge-invariant features of the field everywhere, not just in $X$, as we have seen. Nor are the Lorenz and $R_\xi$ gauges, for similar reasons.

How general is this? It is highly plausible (albeit somewhat heuristic) reasons to expect that a large class of gauge-fixings will fail to be local descriptors. Specifically, consider \emph{local} gauge-fixings, in which the gauge is fixed by requiring
\be\label{localgaugefixing}
\mc{H}(x,A_\mu{x},\partial_\nu A_\mu(x),\ldots \psi(x),\partial_\mu \psi(x), \ldots)=0
\ee
for all $x$ and for some function $\mc{H}$, with the $\ldots$ indicating possible dependence on finitely many higher derivatives. (All gauge-fixings I know are local gauge fixings; it is hard to see how any other gauge-fixing could provide a local descriptor, though I know of no rigorous results here.) Suppose that $(\psi,A)$ is a field and that $\Lambda$ is a gauge transformation which transforms it into a field that satisfies the gauge-fixing condition, and suppose also (which is generically the case) that the gauge group acts freely on $(\psi,A)$, \iec no nontrivial gauge transformation leaves $(\psi,A)$ invariant, so that $\Lambda$ is unique. We will then have some equation for $\Lambda$, of form
\be
\mc{H}(x,A_\mu(x)+\partial_\mu \Lambda(x),\partial_\nu (A_\mu(x)-\partial_\mu \Lambda(x)),\ldots,\psi'(x)\e{ie\Lambda(x)},\partial_\mu (\psi'(x)\e{ie\Lambda(x)}),\ldots)=0.
\ee 
It is clear that in general this will be a \emph{differential} equation, which in general will suffice to uniquely determine $\Lambda$ within $X$ only given boundary conditions (or other global conditions on the solution) on $X$. For the gauge-fixing to be well-defined on $\mc{M}$ we will have to assume that the conditions (\ref{localgaugefixing}) are supplemented by some such conditions on $\mc{M}$. But these do not in general determine boundary conditions on some subset $X\subset \mc{M}$: in general, even if we can solve the equations inwards from the boundary of $\mc{M}$ (which is by no means guaranteed) the boundary data on $X$ will depend not only on the boundary conditions at $\mc{M}$ but also on the fields on the complement of $X$. Hence, two fields that are gauge-equivalent on $X$ will in general not be gauge-fixed into fields that are equal on $X$. 

The only way around this argument is for $\mc{H}$ to depend at most on $\psi(x)$ and $x$, but not on $A(x)$ or any derivatives of $\psi$ or $A$: only in this case will the gauge condition be an \emph{algebraic} constraint on $\Lambda$. Since the modulus of $\psi$ is gauge-invariant, this effectively makes the gauge condition into a condition on the phase of $\psi$: writing $\psi(x)=\rho(x)\e{ie\phi(x)}$, the gauge condition becomes
\be
\phi(x)=f(x,\rho)
\ee
for an arbitrary smooth function $f(x,\rho)$. Call this gauge-fixing condition \emph{generalized unitary gauge} (fairly clearly it will almost always be simpler in practical applications just to use unitary gauge).

\section{The limitations of unitary gauge}\label{limitations}

To sum up: most gauge-fixings provide a gauge-invariant description of a theory, but in most cases that description is highly nonlocal, describing the fields as a whole but providing no reliable insight into the spatial structure of those fields' gauge-invariant features. Unitary gauge (and its generalizations) are an exception, providing a description of the fields that is gauge-invariant, local, and separable.

Furthermore, unitary gauge has a natural interpretation in terms of directly-physical, obviously gauge-invariant features of a theory. The (always-real, always-positive) field $\psi$ in unitary gauge represents the manifestly gauge-invariant quantity $|\psi|$. (In a generalized unitary gauge, this is represented by $\psi\e{-ief(x,|\psi|)}$). The connection $A$ satisfies the equation
\be
\mc{D}_\mu \psi = \partial_\mu \psi +  i e  A_\mu \psi
\ee
and in unitary gauge $\psi$ is real, so that $A_\mu \psi$, up to charge, is the imaginary part of the covariant derivative. If in general we write $\psi=\rho \e{ie\phi}$ and then decompose $\e{-ie\phi}\mc{D}_\mu\psi$ into real and imaginary parts, we obtain
\be
\mc{D}_\mu \phi = \partial_\mu\phi + e A_\mu
\ee
which in unitary gauge reduces to
\be
A_\mu=\frac{1}{e}\mc{D}_\mu \phi 
\ee
so that $A$, in that gauge, just represents the (again, manifestly gauge-invariant) covariant derivative of the phase. (See \cite{struyvegauge} for further consideration of this close relationship between unitary gauge and a manifestly gauge-invariant description.)

This might suggest a preferred status for (generalized) unitary gauge: it, and it alone, provides a local, separable, representation of the gauge-invariant features of a field. There are two reasons to be cautious of this conclusion, which I will refer to as the Degeneracy Reason and the Incompleteness Reason.

The Degeneracy Reason is that unitary gauge is not actually a gauge-fixing whenever $\psi=0$ on an open subset of $\mc{M}$. For then  we can assign the phase of the scalar field arbitrarily, and more importantly, then we cannot recover the connection (and hence the gauge-invariant field strength) from the (vanishing) covariant derivative of $\psi$.

There is something important to this objection (to which I return below) but in itself it should not be taken seriously. If $\psi$ \emph{literally vanishes} on an open set $X$, then unitary gauge is ill-defined on $X$. But if $0<|\psi|<10^{-10^{100}} \mathrm{kg}$ on $X$, unitary gauge is perfectly well defined.\footnote{$|\psi|$ has the dimensions of mass in units where $\hbar$=1. If you prefer SI units, use them; manifestly the point I make here is independent of a choice of scale.} And to suppose that it is physically significant whether $|\psi|=0$ or merely $|\psi|<10^{-10^{100}} \mathrm{kg}$ is to take scalar electrodynamics far more literally than it has any right to be taken. It is an approximately-accurate description of the structure of certain systems on certain scales, not a candidate for the Theory of Everything.

The point is only sharpened in quantum field theory, where classical fields are not direct descriptions of a system but merely variables in a path integral. That path integral requires regularization to be well-defined (the regularization makes the space of fields finite-dimensional so that a measure can be defined straightforwardly), but under any reasonable regularization, fields that vanish on an open set will have measure zero, and so will not contribute to the path integral.

That said, the Degeneracy Reason does suggest that unitary gauge will be misleading or unhelpful --- even if strictly correct --- when applied to regions where the matter field is negligible. One way to characterize the abstract structure of a gauge theory is that a choice of gauge corresponds to a reference frame (for scalar electrodynamics, a choice of the real axis) in the internal field space at each point in space, and that the connection tells us how to relate those reference frames along any given path. Unitary gauge amounts to letting the matter field itself determine the reference frame; this has clear physical significance when the matter field is dynamically relevant, but seems rather arbitrary when it is negligible. 

Similarly, if we are using quantum-field-theoretic methods to study electrodynamics in a region where there are no matter particles, it is natural to integrate out the matter field entirely, obtaining an effective field theory for the electromagnetic field alone. The resulting theory will have only $A$ as its degree of freedom, and so we cannot define unitary gauge. (It remains well-defined in the full quantum field theory, but the degrees of freedom it uses are dynamically irrelevant in this regime.) 

The Incompleteness Reason is more compelling, and applies whether or not the matter field is negligible. Consider a constant phase shift, $\psi(x)\rightarrow \psi\e{ie \theta}$, where $\theta$ is spatially independent. Manifestly, the unitary-gauge description of a field is invariant under that transformation: it leaves both $A$ and $|\psi|$ invariant. But the physical features of the system are \emph{not} invariant, for this transformation is not a small gauge transformation: it is a physical shift of the system, representing a change in its phase as measured by an observer outside the system. This means that, in the terminology of section \ref{introduction}, unitary gauge is not a gauge-fixing at all, since it does not assign an element to every gauge equivalence class: if the phase of $\psi$ tends to anything other than zero at infinity, there is no small gauge transformation which puts $\psi$ in unitary gauge. Put another way: putting a field into unitary gauge discards genuine information --- the average phase of the scalar field --- along with descriptive fluff. If the Degeneracy Reason gives cause to worry that unitary gauge is not a gauge-invariant descriptor, the Incompleteness reason gives cause to worry that it is not a complete descriptor.

(Similarly, if a gauge theory is defined on a topologically non-trivial manifold, the gauge transformation required to put a field in unitary gauge may be topologically large; this occurs in the solenoid spacetime of section~\ref{introduction}. Again, this means that unitary gauge conflates two fields that are distinguished one from another by their relation with external systems, though the nature of that relation is subtler here --- see \cite{wallace-isolated-2}.)

It is helpful here to compare gauge theory to Newtonian mechanics. In the latter theory, it is fairly straightforward \cite{BarbourBertotti1982,barbour2010} to start with the dynamics of an isolated system and then quotient out its translational and rotational symmetries; the result is a self-contained dynamical theory of how the intrinsic features of the theory change with time. But doing so throws away genuine information, since (at least in most applications) Newtonian mechanics describes not a fictional system in an otherwise-empty universe but a \emph{de facto} isolated subsystem of a larger cosmos, and symmetry-related states of that system are differently located and oriented with respect to that larger cosmos~ \cite{wallace-isolated-1}. In the same sense, the unitary-gauge description of a field can helpfully be understood as representing the intrinsic features of an isolated subsystem. A single additional, nonlocal, piece of information --- its average phase --- is required to specify the system completely, at least if it too is understood as an isolated subsystem in a larger cosmos.

But the Incompleteness Reason has an important limitation: it assumes that global phase shifts actually are symmetries of the theory. Whether this is true depends sensitively on the boundary conditions of the theory. If that boundary condition imposes $\lim_{|\vctr{x}|\rightarrow \infty}\psi=0$, it is preserved by global phase shifts, and so $\psi$ and $\psi\e{ie\theta}$, for constant $\theta$, are distinct possible states. But if it imposes a nonzero infinite limit $\psi_0$ for $\psi$, that limit will not be preserved under global phase shifts and so they will not represent a physical symmetry. In the latter, but not the former, case, (generalized) unitary gauge is indeed a gauge-fixing, and moving to (generalized) unitary gauge provides a lossless description of a field's gauge-invariant properties.

The distinction between these two scenarios can be usefully illustrated by considering the spectrum of elementary excitations of the theory, which in classical field theory is the space of solutions of the linearized field equations, and in quantum field theory is the space of particles. In either case we obtain it by looking at the quadratic part of the expansion of the action of the theory around its minimum. In classical field theory the action is given once and for all and determines the classical equations of motion; in quantum field theory the action is the (temperature-dependent) effective action and its successive derivatives determine the vacuum expectation value and the one-particle-irreducible (1PI) $N$-point functions, with the quadratic part of the expansion determining the (inverse of the connected part of the) 2-point function and hence the mass spectrum.

In the case of an unbroken symmetry (boundary conditions where the matter field vanishes) the elementary excitations consist of two excitations of the matter field, related by a 90-degree phase shift (the signature of this in QFT is antimatter; cf \cite{BakerHalvorson2009,wallaceantimatter}) and describable by a complex scalar field, and two excitations of the gauge connection, forming a massless vector field (the exact description of which varies from gauge to gauge). Here the existence of multiple phase-related modes of the matter field is entirely hidden in unitary gauge in the quadratic expansion (of course it can be seen at higher order points in the expansion) while being manifest in Lorenz, Coulomb, or $R_\xi$ gauge. 

The awkwardness of accessing the spectrum from unitary gauge can be traced back to the Degeneracy Reason: we are trying to expand the field around $\phi=0$, but unitary gauge fails on that configuration, and so the expansion is badly behaved.

If the symmetry is broken (in other words, if the boundary conditions are such that the matter field is nonvanishing) we can safely use unitary gauge to study the excitation spectrum; here, as any elementary QFT textbook will tell you, we encounter the Higgs mechanism, where the excitations comprise a single excitation of the matter field describable by a real scalar field, and three excitations of the gauge connection which jointly form a massive vector field (and which are described in an entirely local, gauge-invariant manner if we gauge-fix to unitary gauge).

(What becomes of the analogy to Newtonian mechanics in this case? The short answer is that if the symmetry is spontaneously broken, we cannot simply combine isolated systems with different boundary conditions to make a larger system. For the longer answer, see \cite{wallace-isolated-2}.)

I pause to note two subtleties. Firstly, in classical field theory whether a symmetry is broken or not is a once-and-for-all feature of a theory: it is fixed by the action, and the action encodes the fundamental dynamics. In quantum field theory, by contrast, the action is an effective function determined by the temperature, and the same theory can display different symmetry-breaking phenomena at different temperatures. It thus makes literal sense in QFT (but only metaphorical sense in a classical theory) to talk about symmetry-broken and symmetry-unbroken \emph{sectors} of the same theory.

Secondly, this attribution of excitations to matter rather than gauge connection, or vice versa, should be taken with a grain of salt, since it is not gauge invariant. (We can study the symmetry-broken sector of scalar electrodynamics in $R_\xi$ gauge perfectly well; if we are feeling masochistic we can even study it in Coulomb or Lorenz gauge.) It is of course gauge-invariant to say that \emph{in unitary gauge} the elementary excitations of the symmetry-broken sectors comprise one matter and two gauge-connection excitations.

\section{Locality and unitary gauge in non-Abelian gauge theories}\label{beyond}

(This section lies outside the main flow of the paper, and may be skipped on first reading.)

The general arguments of section~\ref{locality} apply to non-Abelian gauge theories too: in order to be a local descriptor, a gauge-fixing must algebraically determine the gauge transformation required to take any field into its gauge-fixed form, and this means it must fix the gauge-variant features of the matter field. For the moment, consider this problem for a scalar matter field, taking values in some space $\mc{V}$ on which the gauge group $G$ acts. We can partition $\mc{V}$ into \emph{orbits}, equivalence classes under the action of $\mc{V}$; the generalization
of unitary gauge requires us to pick, in each orbit, some representative element.\footnote{We could choose to let the choice of representative element depend on spatial location, but nothing is gained by such generalization.} Let $f:\mc{V}\rightarrow \mc{V}$ map each point in $\mc{V}$ to the representative element in its orbit; then we gauge-fix a field $(\psi,A)$ by applying a gauge transformation $u$ such that $u(x)\psi(x)=f(\psi(x))$. (For scalar electrodynamics, $f$ is given by $f(\psi)=|\psi|$.)

Does this actually determine a gauge-fixing? Only if it fixes $u$ uniquely, which requires that no gauge transformation $u'$ distinct from $u$ exists such that $u(x)\psi(x)=u'(x)\psi(x)$, which is equivalent to the requirement that no non-trivial gauge transformation $v$ exists such that $v(x)\psi(x)=\psi(x)$ (for any $x$). In other words, the group $G$ should act \emph{freely} on every orbit of $G$ in $\mc{V}$ in which $\psi(x)$ takes values. Requiring gauge transformations to be smooth weakens this slightly: there must be no orbit of $G$ in $\mc{V}$ such that (i) $G$ does not act freely on $\mc{V}$; (ii) there is an open set in spacetime such that $\phi$, restricted to that set, takes values in the orbit.

For scalar electrodynamics, there is only one such non-free orbit: the point $\psi=0$, already discussed. But in many scalar non-Abelian theories, there will be many. Take, for instance, the standard action of $SU(N)$ or $U(N)$ on $\Co^N$. The orbits are the (complex) spheres of fixed $|\psi|$, and $SU(N)$ does not act freely on any such orbit: for any nonzero vector in $\Co^N$, there will be a $U(1)$ subgroup of $SU(N)$ that leaves it invariant. (This is essentially the mathematics behind the breakdown of electroweak $SU(2)\times U(1) \equiv U(2)$ symmetry to $U(1)$ symmetry in the Standard Model.)

So unitary gauge is not well-defined for non-Abelian gauge theories with scalar matter, at least in these important cases. But in general it will be definable for more complicated (and realistic) matter fields. If $\psi$ is a spinor matter field, for instance, $\psi(x)$ can be thought of as a linear map from $\Co^2$ to $\mc{V}$, with the gauge group acting on the space of such maps; generic points in the space of such maps transform freely under gauge transformations, so that specifying a representative element in each orbit will fix the gauge of any field except for some non-generic ones. Unitary gauge then works as well for non-Abelian as for Abelian theories: that is, it works, up to the Degeneracy and Incompleteness reasons, which apply to the non-Abelian case \emph{mutatis mutandis}.

It must be admitted, though, that there is something awkward and cumbersome about these generalizations of unitary gauge (and indeed I am not aware of any such generalization having applications in the physics literature). We can understand why this is by considering again the interpretation of unitary gauge I offered in the previous section: a choice of gauge amounts to a choice of reference frame at each point; unitary gauge uses the matter field itself to fix a choice of reference frame. For scalar electrodynamics there is an obvious best way to do so: any nonzero field at a point has a phase and the connection along a path between any two points can be interpreted simply and invariantly as giving the change in phase between the fields at those two points. When the matter field is more complicated, there may be no obviously most natural rule to determine a reference frame or to define the transformation between two such frames, even when the field has sufficient structure that such rules can be stated. (Similar issues would arise even in electrodynamics given a more complicated, e.g. spinor, matter field.)

\section{Conclusion}\label{conclusion}

Gauge-invariant descriptions of a theory with local gauge symmetry are easier to come by than you may think: any gauge-fixing for a theory provides a gauge-invariant description of that theory's fields, just as any description of a spatial system in an antecedently-specified coordinate system provides a coordinate-invariant description of that spatial system.

However, most such descriptions are deceptive. On the surface they appear to be a \emph{local} description of the system (specified as they are in terms of the values of local fields), but in reality they are generally nonlocal, in the sense that two fields gauge-equivalent in a region can often have distinct gauge-fixings in that region.

The exception is unitary gauge (along with  its generalizations). Unitary gauge is local (thus improving on most other gauges), gauge-invariant (thus improving on a description in terms of field strengths) and separable (thus improving on a description in terms of holonomies), and explicitly includes matter degrees of freedom (which field-strength and holonomy descriptions do not). However, in the unbroken sector of a gauge theory, unitary gauge provides an incomplete description of the system (although a complete description of the intrinsic features of the system), since it leaves out overall phase and potentially also topological information; in cases more complicated than scalar electrodynamics it may also be quite awkward, and there may be no natural reason to choose one version of unitary gauge over another.

What does this mean for the search for gauge-invariant descriptions of physical processes? In a sense they are easy to come by: pick a gauge, describe the physics in that gauge, and call it a day. But the nonlocal aspects of most gauges may make the description thus obtained (gauge-invariant, but) somewhat opaque. In many cases we can avoid this problem by using unitary gauge, but this must be done with some caution if asymptotic symmetry is unbroken, since that description is incomplete (although it ought still to be adequate if all that is desired is a description of the system's intrinsic behavior), obscures the particle spectrum, and is awkwardly suited to describing situations where the matter field is negligible.

The Higgs mechanism provides a brief illustration. Philosophers of physics have long\footnote{See, \egc, \cite{earmancurie,smeenkhiggs,lyrehiggs}.} sought a gauge-invariant description of it: here is one (closely related to that offered by \cite{struyvegauge}). Firstly, we must ask whether asymptotic global symmetry is broken in the sector of the theory being studied: this can be characterized simply and gauge-invariantly, and its `tell' is the nonvanishing expectation value of the magnitude of the Higgs field in the ground state (or at thermal equilibrium in a finite-temperature sector). If it is unbroken, the Higgs mechanism does not apply, and the symmetry is manifest in the particle spectrum. If it is broken, unitary gauge offers a complete, gauge-invariant, local, separable description of the physics, in which it is easy to see that the vector boson field is massive and that there is no Goldstone mode.

I close with remarks aimed at a somewhat different audience. I have argued that gauge fixings offer gauge-invariant descriptions of the physics, so that different gauge-fixings offer different gauge-invariant descriptions. But which description is best? The question has an obvious pragmatic interpretation: `best' means `most helpful for the purpose at hand', in which case we should adopt whichever choice we find most helpful (even a nonlocal choice), with little care as to the rightness of that choice beyond its use. 

But there is a tradition in the metaphysics of physics (illustrated very clearly by \cite{sidertools}), where the question is rather: which description is \emph{most fundamental}, which description \emph{carves nature at its joints}? For reasons I explicate in \cite{wallacemathfirst}, I am skeptical that there is a meaningful question here once we set pragmatism aside. For those who disagree, I recommend unitary gauge as the clearly-preferable choice, notwithstanding its limitations when applied to subsystems of the larger cosmos --- an issue which in any case I suspect will not bother those whose metametaphysics is less deflationary than mine. But I also note that the range of possible versions of unitary gauge for systems more complicated than scalar electrodynamics may ultimately cause metaphysical difficulties for any such strategy.

\section*{Acknowledgements}

I am grateful to Philipp Berghofer and Henrique Gomes for useful conversations, and to two anonymous referees for helpful feedback.

%\bibliography{../bib/general2.bib}
%\bibliographystyle{chicago}

\end{document}